\documentclass[journal=jacsat,manuscript=article]{achemso}
\usepackage{physics}
\usepackage[version=3]{mhchem} 
\usepackage{float}
\usepackage{bm}
\usepackage{graphicx}
\usepackage{caption}
\usepackage{subcaption}
\usepackage{amsmath}
\usepackage{listings}
\usepackage{amssymb,mathtools}
\usepackage{soul}
\usepackage{braket}
\usepackage{relsize}
\usepackage{makecell}
\usepackage{placeins}
\usepackage{color,soul}
\usepackage{flexisym}
\usepackage[table]{xcolor}
\usepackage{expl3}
\usepackage{booktabs}
\usepackage{braket}
\usepackage{hyperref}
\usepackage{cleveref}
\usepackage{verbatim}
\usepackage[percent]{overpic}

\author{Shreya Verma}
\affiliation[University of Chicago]
{Department of Chemistry, University of Chicago, Chicago, IL 60637, United States}
\author{Antonio L. Mariano}
\affiliation[Trinity College Dublin]{School of Physics, AMBER and CRANN Institute, Trinity College Dublin, Dublin 2, Ireland}
\email{a.lorenzo.mariano@tcd.ie}
\author{Matthew R. Hermes}
\affiliation[University of Chicago]
{Department of Chemistry, University of Chicago, Chicago, IL 60637, United States}
\author{Alessandro Lunghi}
\affiliation[Trinity College Dublin]{School of Physics, AMBER and CRANN Institute, Trinity College Dublin, Dublin 2, Ireland}
\author{Giulia Galli}
\affiliation[University of Chicago]
{Department of Chemistry, University of Chicago, Chicago, IL 60637, United States}
\alsoaffiliation[University of Chicago]{ Pritzker School of Molecular Engineering, University of Chicago,
Chicago, IL 60637, United States}
\alsoaffiliation[Argonne National Laboratory]{Materials Science Division, Argonne National Laboratory, Lemont, Illinois 60439, United States}
\email{gagalli@uchicago.edu}
\author{Laura Gagliardi}
\affiliation[University of Chicago]
{Department of Chemistry, University of Chicago, Chicago, IL 60637, United States}
\alsoaffiliation[University of Chicago]{ Pritzker School of Molecular Engineering, University of Chicago,
Chicago, IL 60637, United States}
\alsoaffiliation[University of Chicago]{James Franck Institute, University of Chicago,
Chicago, IL 60637, United States}
\email{lgagliardi@uchicago.edu}

\title[An \textsf{achemso} demo]
  {Multireference Density Matrix Embedding for Spin–Phonon Relaxation}

\abbreviations{DMET,SOC}

\begin{document}

\begin{tocentry}
\centering
\includegraphics[width=1\linewidth]{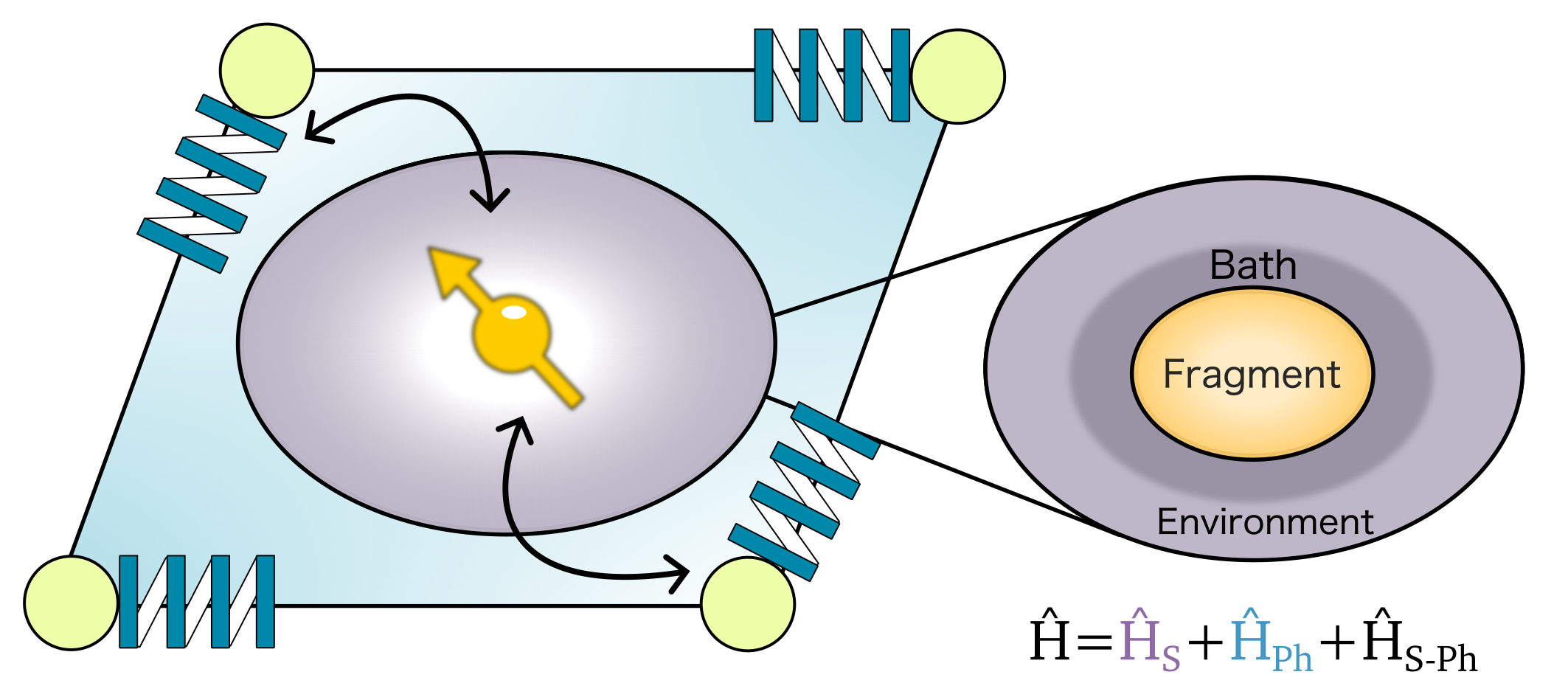}
\end{tocentry}

\begin{abstract}
Spin–phonon coupling governs magnetic relaxation in numerous systems including single-molecule magnets and molecular spin qubits. In most cases, the accurate prediction of spin relaxation rates requires multireference electronic structure methods, but their computational cost has largely restricted such calculations to isolated molecules.
Here we show that spin–phonon relaxation rates can be computed within a multireference density matrix embedding framework. We apply the approach to three cobalt- and two dysprosium-based single-molecule magnets and to a cobalt-based molecular crystal.
Across all systems, treating only the first coordination sphere of the magnetic center at the multireference level reproduces spin relaxation rates in good agreement with non-embedded CASSCF calculations while reducing the correlated problem to 10–68\% of the total basis functions. 
Periodic calculations further demonstrate that spin relaxation rates can be computed for a molecular crystal using an embedded active space of only 259 basis functions out of a total of 2569. These results show that multireference density matrix embedding extends quantitative spin–phonon relaxation calculations from isolated molecules to molecular crystals.
\end{abstract}

\section{Introduction}
The ability to control electronic spin relaxation is central to the development of molecular spin qubits and single-molecule magnets for quantum information science and spintronics~\cite{wolfowicz2021quantum,wasielewski2020exploiting,vaganov2025chemical,sanvito2011molecular}.
Unlike many solid-state qubits, where spin-lattice relaxation times ($T_1$) can greatly exceed coherence times ($T_2$)~\cite{place2021new}, molecular spin systems often exhibit comparatively short ($T_1$) values that can be of the same order of magnitude as ($T_2$)~\cite{atzori2019second}.
In these systems, the coupling between electronic spins and lattice vibrations governs magnetic relaxation over a broad range of temperatures~\cite{ho2018spin}. The accurate prediction of spin-phonon relaxation therefore requires a balanced description of both the electronic structure of the magnetic center and its response to nuclear motion.
This task is particularly challenging in transition-metal and lanthanide complexes, where near-degenerate electronic states and strong electron correlation play a central role in determining magnetic anisotropy and relaxation pathways~\cite{sessoli1993magnetic,rocha2006spin,atzori2019second,vaganov2025chemical}. 

Recent developments have enabled multiscale frameworks for computing spin-phonon relaxation rates in solid-state defects and strongly correlated systems based on Kubo linear response theory \cite{drigo2026spin}, and secular Markovian quantum master equations combined with the \textit{ab initio} calculation of spin-phonon coupling matrices~\cite{lunghi2022computational,lunghi2023spin,nabi2023accurate,kragskow2023spin,xu2020spin}.
In these approaches, thermal lattice vibrations act as a time-dependent perturbation to the spin Hamiltonian, inducing transitions between magnetic states that drive spin relaxation. Depending on the order of perturbation theory in the spin-phonon coupling, both one- and two-phonon relaxation processes can be described within the same formalism. 
Unlike phenomenological models, the quantities entering the relaxation dynamics in this framework are obtained directly from electronic structure and vibrational calculations.
As a result, this fully \textit{ab initio} workflow has enabled quantitative predictions of spin-lattice relaxation rates for a variety of transition-metal and lanthanide complexes~\cite{lunghi2022toward}. Their accuracy, however, depends critically on the underlying electronic structure description~\cite{haldar2025role}.

Multireference methods, such as complete active space self-consistent field (CASSCF)\cite{roos1980complete} and post-CASSCF extensions~\cite{roos1995multiconfigurational,angeli2001introduction,li2014multiconfiguration,wardzala2026multireference,verma2025multireference}, are often required to describe the near-degenerate electronic states, magnetic anisotropy, and spin-orbit-coupled multiplets that govern spin relaxation in transition-metal and lanthanide complexes. Consequently, these methods have become the standard electronic structure approach for \textit{ab initio} spin-phonon calculations~\cite{mondal2022unraveling,haldar2025role,frangoulis2025generating}.

However, the computational cost of these methods grows exponentially/factorially with system size, limiting their direct application to systems with extended ligands or to situations where secondary coordination spheres play an indirect but non-negligible role in the electronic structure of the magnetic center. Moreover, conventional CASSCF calculations are generally not feasible under periodic boundary conditions because of the prohibitively large active spaces required. This limitation is particularly severe for impurities~\cite{seth2025spin} and solid-state defects~\cite{mondal2023spin}, where a clear boundary between the defect and its environment cannot be defined owing to the hybridized orbitals between the defect and the environment. It also extends to molecular systems, which are rarely truly isolated but are instead embedded within crystalline environments containing counterions, solvent molecules, or extended frameworks. 
For example, atoms beyond the first coordination shell of the dysprosium center in \ce{Dyacac} have been shown to contribute significantly to magnetic relaxation~\cite{briganti2021complete}. Similarly, embedding a \ce{Dy(bbpen)Br} molecule within an electrostatic representation of the crystal environment can improve calculated relaxation rates relative to isolated-molecule calculations, depending on factors such as phonon linewidths and $q$-point sampling~\cite{nabi2023accurate}. Electrostatic embedding alone, however, is generally insufficient for covalent crystals~\cite{ingham2025describing}.
Thus, for multireference approaches to be viable for large molecular and condensed-phase systems, scalable electronic structure methods capable of describing strong electron correlation are required. Recently, alternative low-cost approaches for deriving effective crystal-field Hamiltonians have also been proposed. For example, \citet{peng2025accurate} demonstrated that the crystal-field parameters can be extracted from constrained density functional calculations of rotated mean-field states within the low-energy manifold, achieving an accuracy comparable to multireference methods for several lanthanide complexes. While such approaches offer an attractive route to computing crystal-field spectra at mean-field cost, the extension of these ideas to spin-phonon coupling and its derivatives remains largely unexplored.

Quantum embedding methods offer a promising route to overcome these limitations by partitioning a system into an active region treated with a high-level wave function method and an environment treated at a lower level of theory~\cite{sun2016quantum,verma2025multireference}.
Among these approaches, density matrix embedding theory (DMET) has emerged as a powerful wave function-in-wave function embedding framework~\cite{wouters2016practical,wouters2017five}. DMET maps the full system onto a smaller impurity problem consisting of a chosen fragment and bath orbitals constructed from its entanglement with the environment, enabling high-level treatments of strong correlation within a reduced Hilbert space. Recent developments have extended DMET to multireference settings, including complete active space density matrix embedding theory (CAS-DMET)~\cite{pham2018can,pham2019periodic}, $n$-electron valence state perturbation theory (NEVPT2-DMET)~\cite{mitra2022periodic}, and pair-density functional theory~\cite{mitra2023density,verma2025dmepdft}, enabling accurate and scalable electronic structure calculations for strongly correlated molecular and solid-state systems.
In CAS-DMET, the scaling of CASSCF is reduced from $\mathcal{O}(N^3N_{act}^2)$ to $\mathcal{O}(N_{imp}^3N_{act}^2)$. While CAS-DMET has been successfully applied to excitation energies, optical spectra, and electronic properties, its application to magnetic response properties remains limited. Moreover, it has been shown that DMET-based methods provide more accurate electronic properties for molecules compared to truncating the extended ligands, irrespective of the system~\cite{verma2025dmepdft}. Recent studies have demonstrated that CAS-DMET and NEVPT2-DMET can accurately predict zero-field splitting parameters in single-ion magnets~\cite{ai2025density,guan2025density}. Whether these methods can also provide reliable spin-phonon couplings and magnetic relaxation rates remains an open question.

In this work, we develop an embedded multiscale framework for spin-phonon relaxation by using CAS-DMET, state-interaction spin-orbit coupling, and open quantum system theory. To our knowledge, this is the first application of a wave function-in-wave function embedding approach to spin-phonon relaxation. We apply the method to five molecular magnets from the benchmark set of Ref.~\citenum{mondal2022unraveling}, including three cobalt complexes (Complex-I~\cite{fataftah2014mononuclear}, II~\cite{rechkemmer2016four}, III~\cite{pavlov2016polymorphism}) and two dysprosium complexes (Complex-IV~\cite{liu2016stable}, V~\cite{gupta2016air}). By systematically varying the DMET fragment size, we assess the robustness of embedded spin-phonon calculations and show that inclusion of only the first coordination shell is sufficient to reproduce relaxation rates obtained from non-embedded calculations. We further demonstrate the applicability of the approach to a cobalt-based molecular crystal (Crystal-I), where periodic CAS-DMET enables spin-phonon relaxation calculations that would be computationally prohibitive with conventional periodic CASSCF. These results establish practical guidelines for combining multireference embedding methods with spin-phonon relaxation theory and provide a foundation for studying magnetic relaxation in molecular materials and covalent solids.

\begin{figure}
    \centering
    \includegraphics[width=1\linewidth]{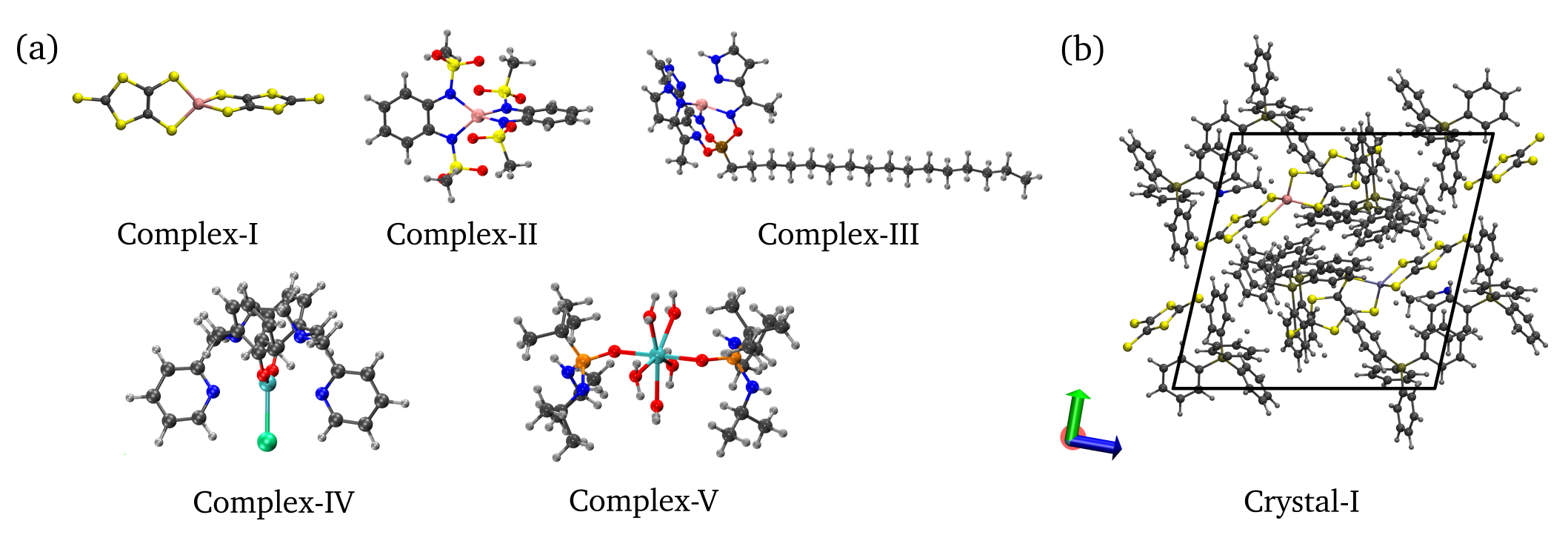}
    \caption{(a) The structures of the five complexes studied with CAS-DMET to compute spin-phonon relaxation. (b) The structure of the optimized conventional unit cell corresponding to Complex-I, containing two molecules of the complex. The atoms are color-coded as follows: Co = pink, Dy = cyan,  Zn = violet, N = dark blue, Cl = green, P = orange, O = red, C = dark gray, and H = light gray.}
    \label{fig:allcomplexes}
\end{figure}

\section{Theory}\label{theory}
\subsection{Density Matrix Embedding Theory} 
Here we outline the key equations of DMET and the embedding Hamiltonian. For a detailed description, we refer the readers to Refs.~\citenum{wouters2016practical,wouters2017five,verma2025multireference}. The system is partitioned into a fragment and its surrounding environment. The wave function $\ket{\Psi}$ of the full system, including the fragment and environment, can be Schmidt decomposed into $n_f$ fragment states, $\ket{\alpha_{f}}$, and an equal number, $n_b=n_f$, of bath states, $\ket{\alpha_{b}}$, entangled to the fragment states:
\begin{equation}
    \ket{\Psi}=\sum_\alpha^{n_f}\lambda_\alpha\ket{\alpha_{f}}\ket{\alpha_{b}}\label{eq:dmet_svd}
\end{equation}
where $\lambda_\alpha$ are the Schmidt coefficients.  
When $\ket{\Psi}$ is a single Slater determinant, this decomposition defines an orbital basis in which the environment is partitioned into a large number of unentangled orbitals which can be treated as a frozen core, and a small number of  bath orbitals entangled with the fragment. The fragment and bath orbitals are collectively referred to as the impurity and the number of impurity orbitals, $N_{\rm{imp}}$,  is at most twice the number of fragment orbitals, $N_f$.
The impurity Schr\"{o}dinger equation can be solved with any quantum chemical methods, here we use CASSCF.

In the periodic extension of CAS-DMET, a gamma-point HF calculation is first performed, after which the localized orbitals are used to define the fragment and construct the corresponding bath and impurity subspace. The embedded Hamiltonian is then solved using CASSCF within the impurity subspace, as in the molecular implementation.

Relativistic effects are included with a scalar relativistic contribution and a spin-orbit coupling (SOC) term. Scalar relativistic effects are introduced with the spin-free exact two-component Hamiltonian~\cite{dyall2001interfacing} and relativistic all-electron basis sets.
Spin-orbit effects are treated perturbatively following a state-averaged CASSCF calculation using the spin-orbit mean field approach~\cite{hess1996mean,jangid2025linearized}. 

\subsection{Spin Hamiltonian with CAS-DMET}
The generalized spin Hamiltonian for the ground $J$-multiplet of the bare magnetic ion of the system in the absence of an external magnetic field, is given by the following form \cite{abragam2012electron} involving tesseral operators ($\hat{O}^l_m$)s and spin Hamiltonian (or crystal field) parameters ($B^l_m$)s:
\begin{equation}
    \hat{{H}}_{S}=\sum_{l=2}^{2J}\sum_{m=-l}^lB^l_m\hat O^l_m\label{eqn:tesseral}
\end{equation}

The spin-orbit Hamiltonian in the basis of the state-averaged CASSCF eigenstates restricted to the ground $J$-multiplet is rotated into the pseudospin basis~\cite{mariano2023spin}. The parameters \(B_m^l\) are then determined by fitting the matrix elements of \Cref{eqn:tesseral} to this Hamiltonian.

Restricting to the ground $J$-multiplet enables the construction of an effective spin Hamiltonian that can be directly related to experimentally measurable magnetic properties. In the context of spin-phonon coupling, this representation is also particularly advantageous because vibronic interactions can be expressed in terms of derivatives of the crystal-field parameters rather than derivatives of the many-electron wave functions~\cite{mariano2023spin}. 


\subsection{Spin-phonon relaxation theory}
Under the assumption of weak-coupling between the spins and the phonon bath, the spin Hamiltonian is expanded as a Taylor series about the equilibrium geometry and truncated at first order. This first order term gives the linear spin-phonon coupling Hamiltonian \cite{lunghi2019phonons,lunghi2020multiple} ($\hat{H}_{S-ph}$): 
\begin{equation}
    \hat{H}_{S-ph}(t)=\sum_{\alpha}\left(\frac{\partial \hat{H}_S}{\partial Q_{\alpha}}\right)_0\hat Q_{\alpha}(t)
    \label{eqn:hsph}
\end{equation}

where $\hat Q_{\alpha}$ represents the $\alpha$-phonon at gamma-point and vibrational frequency $\hbar\omega_{\alpha}$. Considering the effective spin Hamiltonian expressed in terms of crystal field parameters and tesseral operators (\Cref{eqn:tesseral}), the derivative in the linear coupling term above is evaluated using:
\begin{equation}
    \frac{\partial \hat{H}_S}{\partial Q_{\alpha}}=\sum_{\alpha}\sum_{ml}\frac{\partial B^l_m}{\partial Q_{\alpha}}\hat O^l_m\label{eqn:cfpds}
\end{equation}
From time-dependent perturbation theory, treating the spin as a system in a phonon reservoir, the second and fourth order quantum master equations \cite{lunghi2022toward} for the time evolution of the full system (spanning Hilbert spaces of both spin and phonons) density matrix ($\varrho$) are given by:
\begin{subequations}
\begin{align}
    \frac{d\hat\varrho^S(t)}{dt}&=-\frac{i}{\hbar}[\hat{H}_{S-ph}(t)~,~\hat\varrho(0)]-\frac{1}{\hbar^2}\int_0^tds~[\hat{H}_{S-ph}(t)~,~[\hat{H}_{S-ph}(s)~,~\hat\varrho(s)]]\label{eqn:r2}\\
    \frac{d\hat\varrho^S(t)}{dt}&=\frac{1}{\hbar^4}\int_0^tds\int_0^sds'\int_0^{s'}ds''~Tr_{ph}~[H_{S-ph(t)},~[\hat{H}_{S-ph}(s)~,[\hat{H}_{S-ph}(s'),~[\hat{H}_{S-ph}(s'')~,~\hat\varrho(s'')]]]]
    \label{eqn:r4}
\end{align}
\end{subequations}  
where $\varrho^S$ is the spin-reduced density matrix.

The single-phonon transition rate ($\Gamma^{1-ph}$) between two spin states, $\ket{a}$ and $\ket{b}$, is obtained by inserting the linear coupling term of \Cref{eqn:hsph} into \Cref{eqn:r2}:
\begin{equation}
    \Gamma^{1-ph}_{ba}=\frac{2\pi}{\hbar^2}\sum_\alpha\left|\langle b|\frac{\partial\hat H_S}{\partial Q_{\alpha}}|a\rangle\right|^2G^{1-ph}(\omega_{ba},\omega_{\alpha})\label{orbach-rate}
\end{equation}
where 
\begin{align}
G^{1-ph}&=\delta(\omega-\omega_{\alpha})\tilde{n}_{\alpha}+\delta(\omega+\omega_{\alpha})(\tilde{n}_{\alpha}+1)\\
\tilde{n}_{\alpha}&=\frac{1}{e^{\hbar\omega_\alpha/k_BT}-1}\nonumber
\end{align}
The relaxation mechanism involving one-phonon transitions as described by \Cref{orbach-rate}, leading to a total flip in spin orientation is termed as Orbach relaxation.
 
The two-phonon process, namely the Raman relaxation, occurs with the spin transition mediated by intermediate excited spin states, $\ket{k}$. Previous studies have provided substantial evidence that this mechanism can successfully account for experimentally observed Raman relaxation behavior in molecular spin systems~\cite{lunghi2022toward}. The corresponding transition rate ($\Gamma^{2-ph}_{ba}$) between the spin states $\ket{a}$ and $\ket{b}$ accounts for three possible two-phonon processes: absorption of two phonons, emission of two phonons, and simultaneous emission of one phonon and absorption of a second one. The rate is given by:
\begin{align}
    &\Gamma^{2-ph}_{ba}=\frac{2\pi}{\hbar^2}\sum_{\alpha\beta}\left| \sum_{k}\left[\frac{\langle a|\frac{\partial \hat H_S}{\partial Q_\alpha}|k\rangle\langle k|\frac{\partial \hat H_S}{\partial Q_\beta}|b\rangle}{E_k-E_b+\hbar\omega_\beta}+ \frac{\langle a|\frac{\partial \hat H_S}{\partial Q_\alpha}|k\rangle\langle k|\frac{\partial \hat H_S}{\partial Q_\beta}|b\rangle}{E_k-E_b-\hbar\omega_\alpha}\right]\right|^2G^{2-ph}(\omega_{ba},\omega_{\alpha}, \omega_{\beta})\label{raman-rate}\\ 
    &G^{2-ph}(\omega_{ba},\omega_{\alpha},\omega_{\beta\mathbf{q'}})=\delta(\omega_{ba}-\omega_{\alpha}+\omega_{\beta})~\tilde{n}_{\alpha}(\tilde{n}_{\beta}+1)\nonumber
\end{align}
 
Finally, the total relaxation time $\tau=1/\Gamma$ can be calculated using:
\begin{equation}
    \tau = \left(\frac{1}{\tau^{1-ph}}+\frac{1}{\tau^{2-ph}}\right)^{-1}\equiv T_1\label{eq:rates}
\end{equation}

\section{Computational Methods}
\paragraph{Electronic structure:}Complexes I-III and crystal-I contain a Co(II) ion with a $S=3/2$ quartet ground state. All related CASSCF and CAS-DMET calculations were performed with an active space of seven electrons in the five 3d orbitals (7e,5o), with a state average over all ten quartet states, as used in~\citenum{haldar2025role}. For Dy(III)-based complexes IV-V, the minimal active space of (9e,7o) including all electrons in the \ce{4f} sub-shell was employed with a state average over all 21 sextet states~\citenum{haldar2025role}. In all cases, equal weights were used for state average. The CAS-DMET calculations (across $6N$, $N$ being the number of atoms in a system,  geometries for all five complexes) used as starting orbitals meta-L{o}wdin localized orbitals. 

Crystal-I consists of two units of Complex-I in a tetraphenyl phosphonium and acetonitrile environment inside the unit cell (\Cref{fig:allcomplexes}(b)). For the CAS-DMET calculations, we consider the first coordination shell of one of the Co units as the DMET fragment, and in the second unit the Co atom is replaced by a Zn atom for magnetic dilution. Only the displacements of 17 atoms corresponding to the Co-containing unit were used for numerical derivatives of crystal field parameters. 
Again, meta-L{o}wdin localized orbitals were used for bath construction.
The DKH1 Hamiltonian was used for all systems for the spin-orbit coupling treatment with the molecular mean field method.

For Complexes-I and II, we used CC-PVTZ-DK \cite{de2001parallel} basis set for all atoms, while for Complex-III and Crystal-I, CC-PVTZ-DK basis set was used only for Co and first coordination shell N atoms and CC-PVDZ-DK was used for all other atoms. For Complexes-IV and V, ANO-RCC-VTZP was used for Dy atom and ANO-RCC-VDZP for remaining atoms.

\paragraph{Phonons} The optimized cell parameters, geometry and $\Gamma$-point phonons were taken from Ref. \citenum{mondal2022unraveling}. These calculations were performed with periodic density functional theory using the software CP2K~\cite{kuhne2020cp2k}. Cell optimization was performed employing a convergence criterion of $10^{-7}$ au and SCF convergence criterion of $10^{-10}$ au for the energy. A plane wave cutoff of 1000 Ry, DZVP-MOLOPT Gaussian basis sets \cite{vandevondele2007gaussian}, and Goedecker-Teter-Hutter pseudopotentials \cite{goedecker1996separable} were employed for all atoms. The Perdew-Burke-Ernzerhof (PBE) functional and DFT-D3 dispersion corrections were used.

\paragraph{Spin-phonon coupling matrix:}The derivatives of the spin Hamiltonian are first computed with respect to cartesian displacements ($R_i$) using the finite difference method. We use displacements of $0.01$\AA~and $0.02$\AA~to fit the derivatives for complexes, and for the crystal, we use only $0.01$\AA. These numerical derivatives are then transformed to the derivatives with respect to phonon modes using:
\begin{equation}
    \frac{\partial \hat H_S}{\partial Q_\alpha}=\sum_i^{3N}\sqrt{\frac{\hbar}{2\omega_\alpha m_i}}L_{i\alpha}\frac{\partial \hat H_S}{\partial R_i}
\end{equation}
where N is the number of atoms in the unit cell. Since the systems we are studying are well known for two-phonon relaxation through the involvement of virtual spin states, only the first order derivatives of the the spin Hamiltonian (\Cref{eqn:cfpds}) are required for both 1-phonon and 2-phonon transition rates, as explained in the theory section. The transformation of spin Hamiltonian derivatives along with the spin dynamics simulation to obtain the transition rates are performed using the MolForge software~\cite{molforge-code,lunghi2022toward}. The code used for interfacing the electronic structure and spin dynamics simulations is available at \citenum{spinon}.

The complete multiscale workflow used to compute spin–lattice relaxation rates is illustrated in Figure 2. First, the spin-orbit Hamiltonian is obtained using CAS-DMET as described in Section 2. This Hamiltonian is mapped to the generalized spin Hamiltonian to obtain crystal field parameters. In parallel, phonon modes are computed at the periodic density functional theory level. These two components are then used to evaluate derivatives of the spin Hamiltonian with respect to phonon coordinates, yielding the spin–phonon coupling Hamiltonian. Finally, the resulting Hamiltonian is propagated using perturbative or Redfield formalism to obtain the rates, as defined in \Cref{orbach-rate,raman-rate}.

\begin{figure}[H]
    \centering
    \includegraphics[width=1\linewidth]{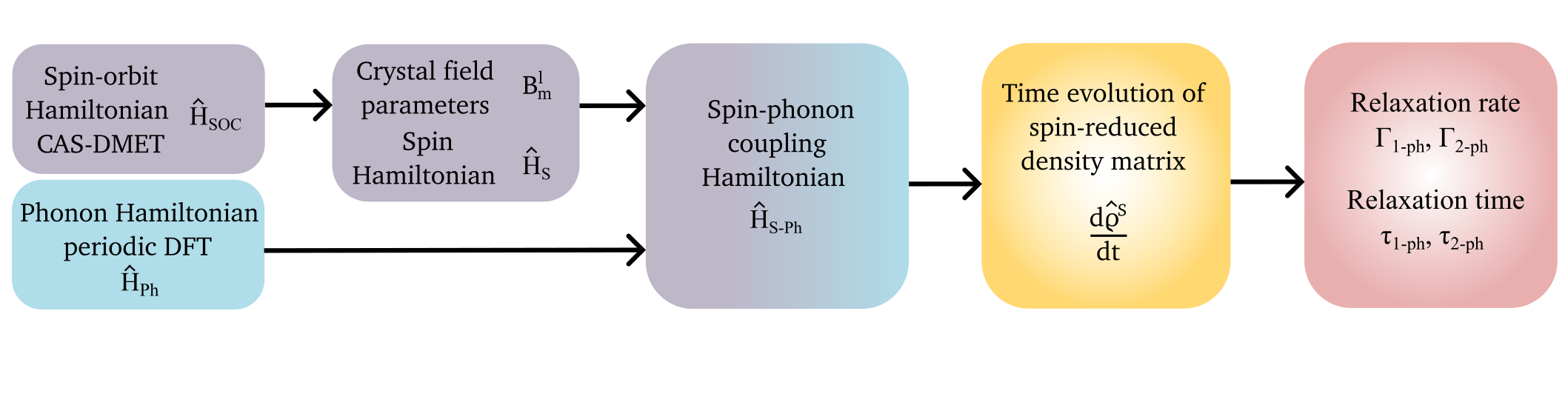}
    \caption{The complete workflow of the procedure to perform spin-lattice dynamics computations. The first step is to compute the spin Hamiltonian, which is done in this work by using density matrix embedding with spin-orbit coupling on top of the embedded CASSCF wave function. On the side, phonons need to be computed at the periodic DFT level. These two ingredients are then used to compute the derivatives of the spin Hamiltonian with respect to the phonon degrees of freedom, which builds the spin-phonon coupling Hamiltonian. The last step is to propagate this Hamiltonian using the perturbation equations or the Redfield equations to finally extract the rates as given by \Cref{orbach-rate,raman-rate}.}
    \label{fig:workflow}
\end{figure}

\section{Results and discussion}
The performance of CAS-DMET for computing spin-phonon relaxation rates is evaluated for five benchmark systems: three cobalt complexes (Complex-I--III), two dysprosium complexes (Complex-IV--V), and a cobalt molecular crystal, studied under periodic boundary conditions (Crystal-I). These systems span diverse coordination environments and provide a test of the robustness of the embedding framework.

\begin{figure}
    \centering
    \includegraphics[width=1\linewidth]{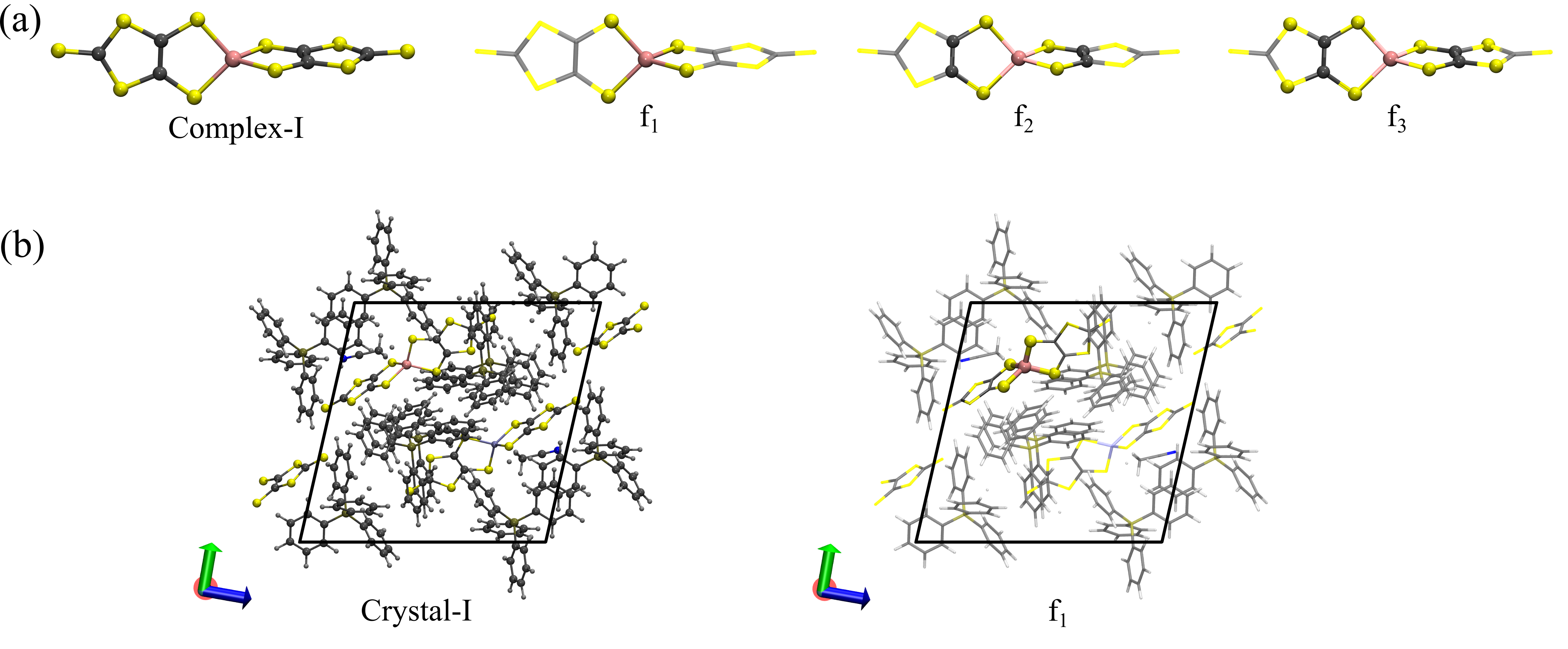}
    \caption{Fragmentation scheme for (a) Complex-I: the full structure, $f_1$, $f_2$ and $f_3$ denoting first, second and third coordination shells as a part of the DMET fragment; (b) Crystal-I: the full structure, $f_1$ fragment considered for the periodic-DMET calculations.  The atoms represented with balls and sticks in $f_1$, $f_2$ and $f_3$ are those included in the DMET fragment, while the environment atoms are represented only by sticks.}
    \label{fig:frag_scheme}
\end{figure}

For each molecular complex, three fragmentation schemes are considered: $f_1$, $f_2$, and $f_3$, corresponding to fragments containing the magnetic center and its first, second, and third coordination shells, respectively. The representative fragmentations for Complex-I are shown in~\Cref{fig:frag_scheme}(a). The resulting impurity sizes are summarized in Table~\ref{tab:nemb}. For Crystal-I, only the $f_1$ fragment was considered (\Cref{fig:frag_scheme}(b)), comprising 259 impurity orbitals out of a total of 2569 basis functions.

\begin{table}[h!]
    \centering
    \begin{tabular}{ccccc}
    \hline
        System & $N_{\mathrm{tot}}$ & $N_{\mathrm{imp}}^{f_1}$ & $N_{\mathrm{imp}}^{f_2}$ & $N_{\mathrm{imp}}^{f_3}$\\
        \hline
        Complex-I & 588 & 316& 433& 561 \\
        Complex-II & 1324 & 340 & 460 & 476\\
        Complex-III & 1035 & 392 & 565 & 653\\
        Complex-IV & 479& 327& 428 & 479\\
        Complex-V & 529 & 335 & 371 & 424\\
        Crystal-I & 2569 & 259 &-&-\\
        \hline
    \end{tabular}
    \caption{Total number of basis functions and impurity orbitals for all embedding fragments used in the CAS-DMET calculations for complexes I–V, and Crystal-I. }
    \label{tab:nemb}
\end{table}

\subsection{Fragment dependence of spin-orbit Kramers' doublets}
\begin{figure}[h!]
    \centering
    \includegraphics[width=1\linewidth]{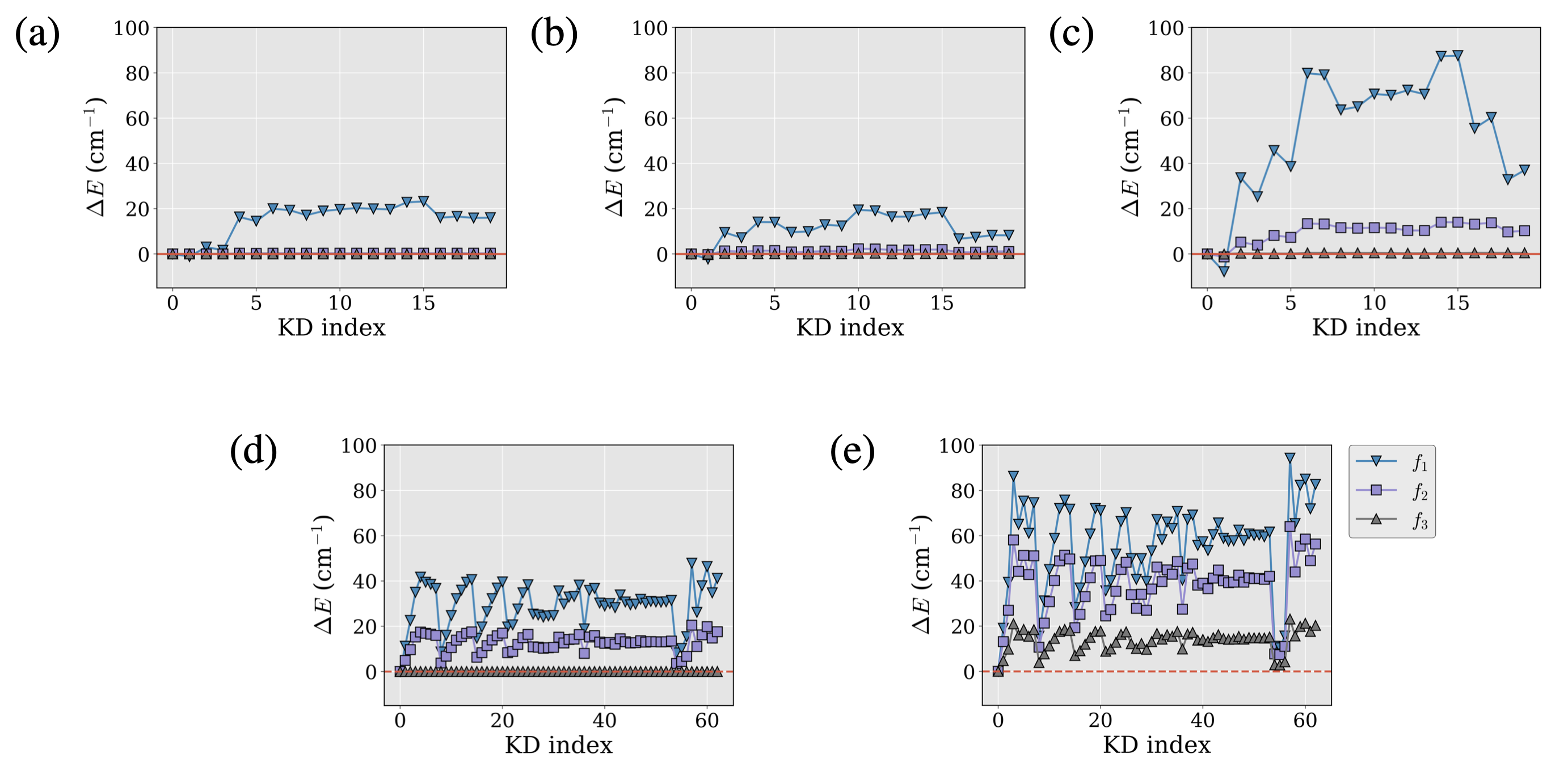}
    \caption{The difference between the Kramers' doublets (KD) energies for each DMET fragmentation ($f_1$, $f_2$ and $f_3$) and their non-embedded CASSCF counterparts are plotted with the KD index for (a) Complex-I, (b) Complex-II, (c) Complex-III, (d) Complex-IV, and (e) Complex-V. The energies are for the equilibrium geometry of each complex.}
    \label{fig:KD_all}
\end{figure}

We first examine the convergence of the spin-orbit Kramers doublet (KD) energies with respect to the DMET fragment size. 
The ground-state KDs are discussed separately for each complex. Here, we focus on the deviations of the CAS-DMET KD energies from the non-embedded CASSCF reference at the equilibrium geometry. \Cref{fig:KD_all} summarizes these deviations for all five complexes, the relative energies are reported in the Supporting Information.
Complexes-I, II, III contain Co(II) centers with a $S=3/2$ ground state resulting in two ground state KDs. 
The ten quartet states included in the state-averaged CASSCF calculation generate twenty KDs. Complexes IV and V contain Dy(III) centers with a $J=15/2$ ground state. The twenty-one sextet states used in the state-averaged calculation give rise to sixty-three KDs.

For the Co-based complexes, the smallest fragment, $f_1$, reproduces the low-lying KDs well. In Complexes I and II, deviations remain below $10~\mathrm{cm}^{-1}$ for the first few excited KDs and increase to approximately $5$-$25~\mathrm{cm}^{-1}$ for higher-energy states. Complex III exhibits larger deviations, reaching $20$-$90~\mathrm{cm}^{-1}$ for the higher excited KDs. A similar trend is observed for the Dy complexes, where $f_1$ deviations increase from approximately $10$-$50~\mathrm{cm}^{-1}$ in Complex IV to $15$-$100~\mathrm{cm}^{-1}$ in Complex V.

Increasing the fragment size systematically improves the agreement with the non-embedded reference. For Complex I, where the $f_2$ fragment already contains approximately 74\% of the basis functions, the differences become negligible. In Complexes II and III, deviations are reduced to below $4~\mathrm{cm}^{-1}$ and $20~\mathrm{cm}^{-1}$, respectively. The Dy complexes converge more slowly, particularly Complex V, where deviations of up to $60~\mathrm{cm}^{-1}$ remain at the $f_2$ level. The largest fragment, $f_3$, yields KD energies that are nearly indistinguishable from the non-embedded results for all systems. The only notable exception is Complex V, where deviations of up to $20~\mathrm{cm}^{-1}$ persist despite the fragment containing approximately 80\% of the total basis functions.

The sensitivity of the higher-energy KDs to fragment size is particularly relevant for two-phonon relaxation processes. While the low-lying KDs are already well described by relatively small fragments, Raman process involve virtual excitations through the full manifold of spin-orbit states (\Cref{raman-rate}). Consequently, inaccuracies in the higher excited KDs can propagate into the calculated relaxation rates, making convergence of the entire KD manifold an important consideration.

\subsection{Cobalt complexes}
\paragraph{Complex-I:} $[{\rm{Co}(\rm{C}_3\rm{S}_5)_2}]^{2-}$ serves as a representative system for assessing how the quality of the embedded electronic structure propagates into spin-phonon relaxation properties. The three fragments considered are $f_1$ ($\rm{Co}(\rm{S}_2)_2$), $f_2$ ($\rm{Co}(\rm{S}_2\rm{C}_2)_2$), and $f_3$ ($\rm{Co}(\rm{S}_2\rm{C}_2\rm{S}_2)_2$).

\begin{table}
    \centering
    \begin{tabular}{ccccc}
    \hline
         States& CASSCF & $f_1$ & $f_2$ & $f_3$\\
         \hline
        $KD_0$ & 0 & 0 & 0 & 0\\
        $KD_1$ & 279 & 277 & 276 & 279\\
        \hline
    \end{tabular}
    \caption{Energies of the lowest Kramers' doublets (in $\mathrm{cm}^{-1}$ for Complex-I using CASSCF and various CAS-DMET fragmentations.)}
    \label{tab:lowestkd1}
\end{table}

\begin{figure}[h!]
    \centering
    \includegraphics[width=1\linewidth]{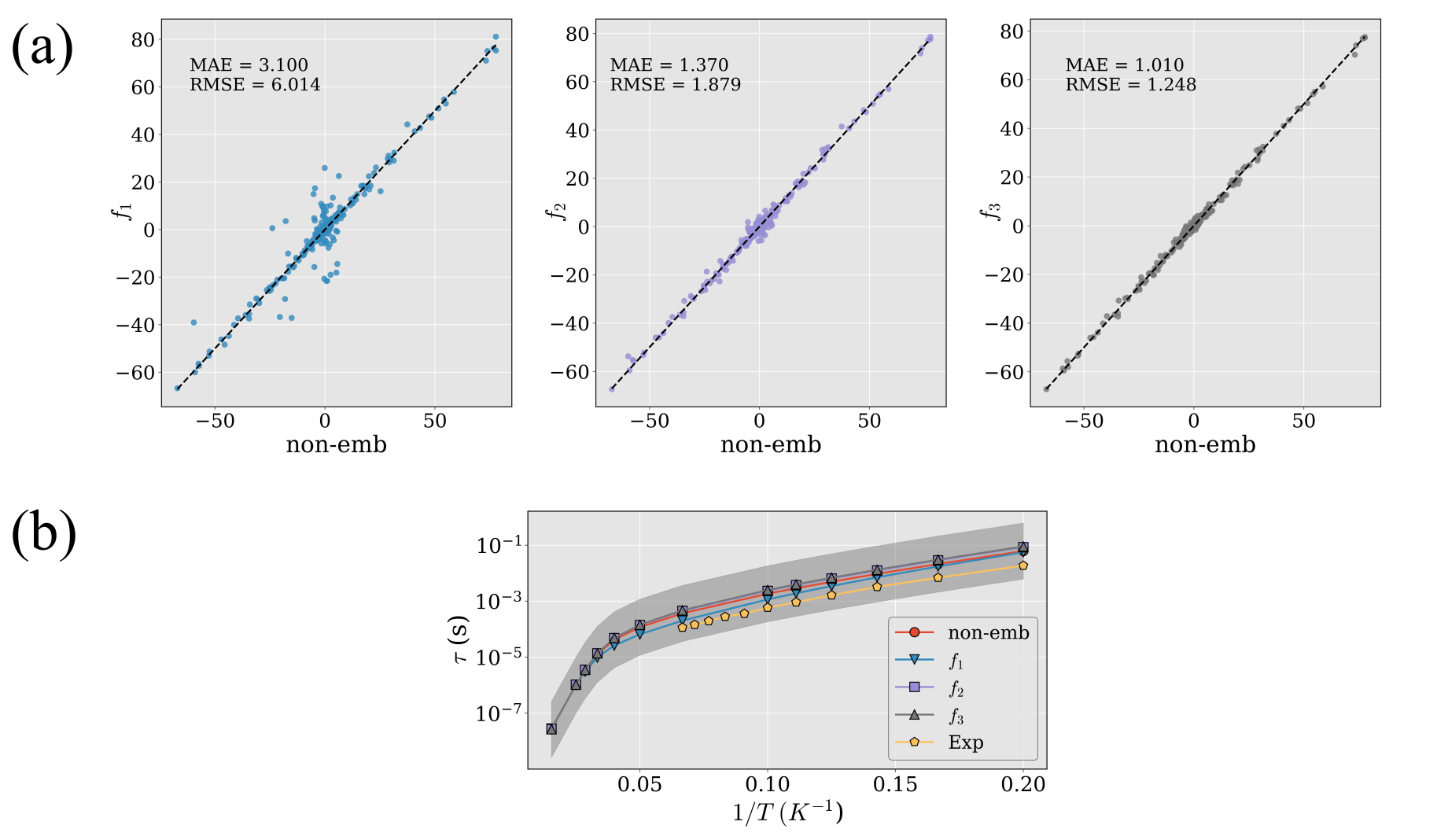}
    \caption{(a) Comparison between the computed crystal field parameter derivatives, $\partial B^l_m/\partial R_i$ ($\mathrm{cm}^{-1}$\AA$^{-1}$), obtained for three DMET fragments $f_1,f_2,f_3$ against the non-embedded CASSCF, respectively, for Complex I. (b) The total relaxation times ($\tau$) (including both Orbach and Raman) in seconds as a function of temperature $(T)$ for Complex I. The dark gray region represents accuracy region of an order of magnitude around the pristine CASSCF relaxation times. The experimental data is taken from \citenum{fataftah2014mononuclear}.}
    \label{fig:co-1-dmet}
\end{figure}

The lowest KD energies are reproduced accurately by all fragmentations (Table~\ref{tab:lowestkd1}), consistent with the trends discussed in the previous section. The corresponding relaxation times are shown in \Cref{fig:co-1-dmet}(b). Even the smallest fragment captures the overall temperature dependence of the non-embedded calculation, while $f_2$ and $f_3$ yield relaxation profiles that are nearly indistinguishable from the CASSCF reference.

The remaining differences can be traced to the crystal-field parameter (CFP) derivatives shown in \Cref{fig:co-1-dmet}(a). The $f_1$ fragment exhibits a broader scatter around the parity line, particularly for CFP derivatives with small magnitudes, whereas increasing the fragment size systematically improves the agreement. The RMSE decreases from $6.014~\mathrm{cm}^{-1}$\AA$^{-1}$ for $f_1$ to $1.248~~\mathrm{cm}^{-1}$\AA$^{-1}$ for $f_3$, indicating smooth convergence toward the full-system limit. Importantly, the residual discrepancies are concentrated among CFP derivatives close to zero, where small numerical differences can produce large relative errors. 
Overall, the parity analysis directly explains the origin of the differences in the spin–phonon relaxation times and further shows that even the $f_1$ fragment reproduces the relaxation times within less than one order of magnitude over the entire temperature range.

\paragraph{Complex-II:} This complex contains a tetrahedral Co(II) center coordinated by four nitrogen atoms and embedded within a larger ligand environment than Complex-I. The three fragments considered are $f_1$ ($\rm{CoN}_4$), $f_2$ ($\rm{Co}(\rm{N}_2\rm{S}_2\rm{C}_2)_2$), and $f_3$ ($\rm{Co}(\rm{N}_2(\rm{SO}_2)\rm{C}_4\rm{H}_2)_2$).

\begin{figure}[h!]
    \centering
    \includegraphics[width=1\linewidth]{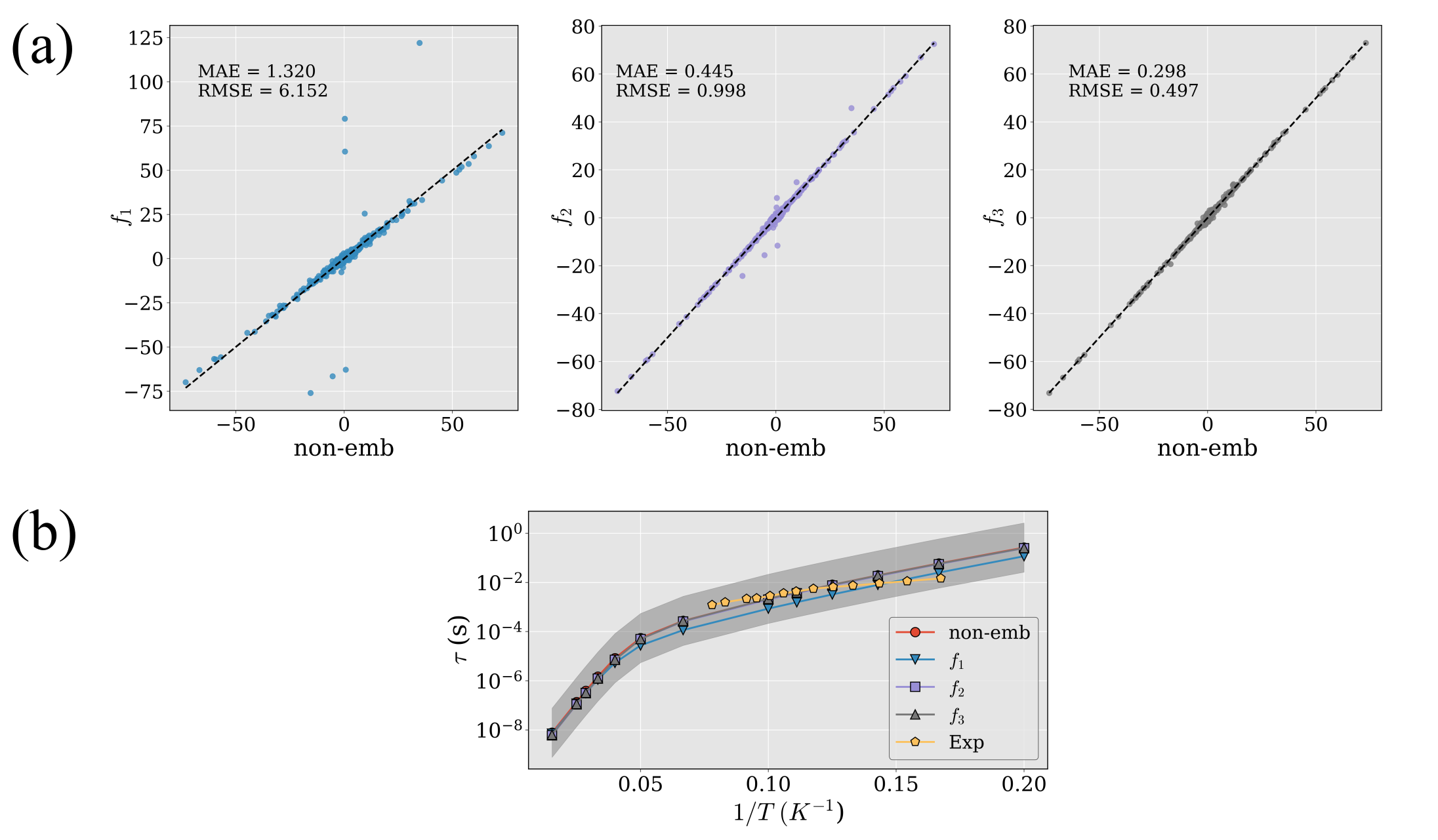}
    \caption{(a) The parity plots for the crystal field parameter $\partial B^l_m/\partial R_i$ ($\mathrm{cm}^{-1}$\AA$^{-1}$), derivatives three DMET fragments $f_1,f_2,f_3$ against the non-embedded CASSCF, for Complex II. (b) The total relaxation times ($\tau$) as a function of temperature $(T)$ for Complex II. The dark gray region represents accuracy region of an order of magnitude around the pristine CASSCF relaxation times. The experimental data is obtained from Ref.~ \citenum{rechkemmer2016four}.}
    \label{fig:co-2-dmet}
\end{figure}

The lowest excited KD is underestimated by approximately $8~\mathrm{cm}^{-1}$ in $f_1$, while $f_2$ and $f_3$ agree with the CASSCF reference to within $1~\mathrm{cm}^{-1}$ (Table~\ref{tab:lowestkd2}). Despite this difference, all fragmentations produce relaxation times that closely follow the non-embedded results (\Cref{fig:co-2-dmet}(b)). The largest deviations occur at low temperatures, where the $f_1$ relaxation times are faster than the reference values, although still within one order of magnitude. Agreement improves further in the high-temperature Orbach regime.

The parity plots in \Cref{fig:co-2-dmet}(a) show that the CFP derivatives are generally well reproduced even for $f_1$, with only a few outliers contributing to the RMSE. Increasing the fragment size further improves the agreement, reducing the RMSE from $6.152~\mathrm{cm}^{-1}$\AA$^{-1}$ for $f_1$ to $0.998~\mathrm{cm}^{-1}$\AA$^{-1}$ and $0.497~\mathrm{cm}^{-1}$\AA$^{-1}$ for $f_2$ and $f_3$, respectively. The residual differences in the $f_1$ relaxation rates can therefore be attributed to a combination of small errors in both the CFP derivatives and the excited KD energies (plot b in \Cref{fig:KD_all}). Additionally, all the simulated rates agree well with the experimental rates.

\begin{table}
    \centering
    \begin{tabular}{ccccc}
    \hline
         States& CASSCF & $f_1$ & $f_2$ & $f_3$\\
         \hline
        $KD_0$ & 0 & 0 & 0 & 0\\
        $KD_1$ & 207 & 199 & 206 & 207\\
        \hline
    \end{tabular}
    \caption{Energies of the lowest Kramers' doublets (in $\mathrm{cm}^{-1}$ for Complex-II using CASSCF and various CAS-DMET fragmentations.)}
    \label{tab:lowestkd2}
\end{table}

\paragraph{Complex-III:} Complex-III features a hexacoordinated Co(II) center surrounded by an extended ligand framework. The long alkyl chain makes this system a useful test of whether regions far from the magnetic center can be treated at the mean-field level without compromising the spin-phonon relaxation dynamics. The three fragments considered are $f_1$ ($\rm{CoN}_6$), $f_2$ ($\rm{CoN}_6\rm{N}_4\rm{O}_2$), and $f_3$ ($\rm{CoN}_6\rm{N}_4\rm{O}_2\rm{C}_6\rm{B}$).

The lowest KD splitting is already well reproduced by the smallest fragment, with $f_1$ differing from the CASSCF reference by only $2~\mathrm{cm}^{-1}$, while $f_2$ and $f_3$ are essentially converged (Table~\ref{tab:lowestkd3}). As discussed in the previous section, the larger differences for this system arise in the higher excited KDs, particularly for $f_1$.

\begin{table}
    \centering
    \begin{tabular}{ccccc}
    \hline
         States& CASSCF & $f_1$ & $f_2$ & $f_3$\\
         \hline
        $KD_0$ & 0 & 0 & 0 & 0\\
        $KD_1$ & 206 & 204 & 206 & 206\\
        \hline
    \end{tabular}
    \caption{Energies of the lowest Kramers' doublets (in $\mathrm{cm}^{-1}$) for Complex-III using CASSCF and various CAS-DMET fragmentations.}
    \label{tab:lowestkd3}
\end{table}

\begin{figure}[h!]
    \centering
    \includegraphics[width=1\linewidth]{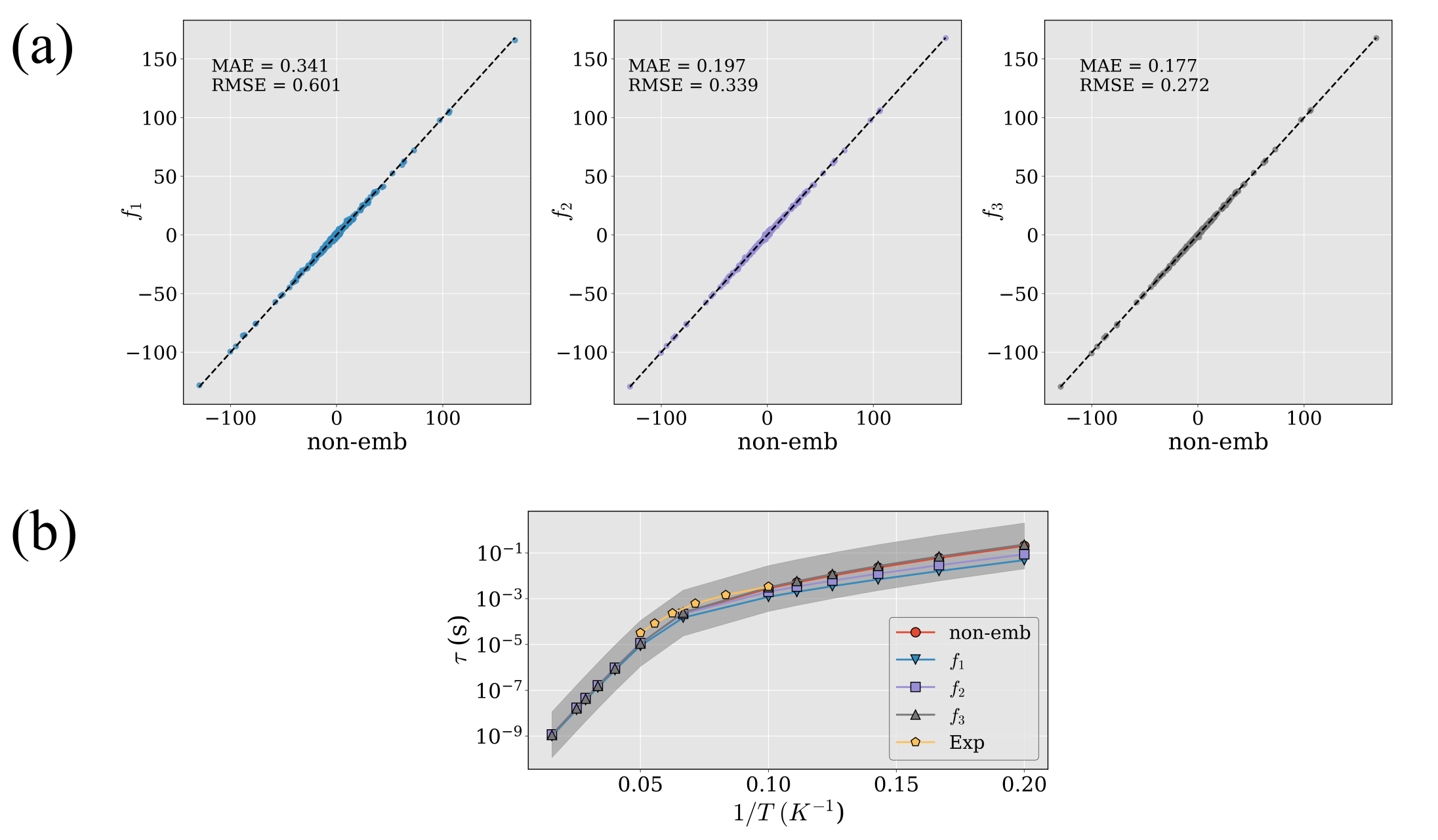}
    \caption{(a) The parity plots for the crystal field parameter $\partial B^l_m/\partial R_i$ ($\mathrm{cm}^{-1}$\AA$^{-1}$), derivatives three DMET fragments $f_1,f_2,f_3$ against the non-embedded CASSCF, for Complex III. (b) The total relaxation times ($\tau$) as a function of temperature $(T)$ for Complex II. The dark gray region represents accuracy region of an order of magnitude around the pristine CASSCF relaxation times. The experimental data is obtained from Ref. \citenum{pavlov2016polymorphism}.}
    \label{fig:co-3-dmet}
\end{figure}

The CFP derivatives are reproduced by all fragmentations (\Cref{fig:co-3-dmet}(a)), with RMSE values below $1~\mathrm{cm}^{-1}$\AA$^{-1}$ in every case. Despite this agreement, the relaxation times obtained with $f_1$ are systematically faster than the non-embedded CASSCF results, especially in the temperature regime where Raman relaxation dominates (\Cref{fig:co-3-dmet}(b)). Increasing the fragment size improves the agreement, with $f_3$ yielding relaxation times that are nearly indistinguishable from the CASSCF reference.

Because both the lowest KD splitting and the CFP derivatives are already well converged for $f_1$, the remaining differences in the relaxation times can be attributed primarily to the larger deviations in the higher excited KDs. This observation is consistent with the discussion above that Raman relaxation depends on virtual excitations through the full spin-orbit manifold. Nevertheless, even the smallest fragment reproduces the relaxation times within an order of magnitude across the full temperature range, demonstrating that the dominant physics remains well captured despite the reduced embedding space.

\subsection{Dysprosium Complexes}
\paragraph{Complex-IV:} Complex IV is a tri-coordinate Dy(III) complex containing the hexadentate \ce{H2bbpen} ligand. Three fragments were considered, corresponding to the first ($f_1$), second ($f_2$), and third ($f_3$) coordination shells of the metal center.
The energies of the eight ground-state KDs arising from the $J=15/2$ manifold are reported in \Cref{tab:lowestkd4}. The first excited KD differs by approximately $11~\mathrm{cm}^{-1}$ in the $f_1$ calculation relative to the CASSCF reference, and the agreement improves as the fragment size is increased.

\begin{table}[H]
    \centering
    \begin{tabular}{ccccc}
    \hline
         States& CASSCF & $f_1$ & $f_2$ & $f_3$\\
         \hline
        $KD_0$ & 0 & 0 & 0 & 0\\
        $KD_1$ & 384 & 395 & 389 & 384\\
        $KD_2$ & 611 & 633 & 620 & 611\\
        $KD_3$ & 701 & 736 & 716 & 701\\
        $KD_4$ & 712 & 753 & 729 & 712\\
        $KD_5$ & 743 & 783 & 760 & 743\\
        $KD_6$ & 781 & 820 & 798 & 781\\
        $KD_7$ & 828 &  865& 844 & 827\\
        \hline
    \end{tabular}
    \caption{Energies of the lowest Kramers' doublets (in $\mathrm{cm}^{-1}$) for Complex-IV using CASSCF and various CAS-DMET fragmentations.}
    \label{tab:lowestkd4}
\end{table}

\begin{figure}[H]
    \includegraphics[width=0.9\linewidth]{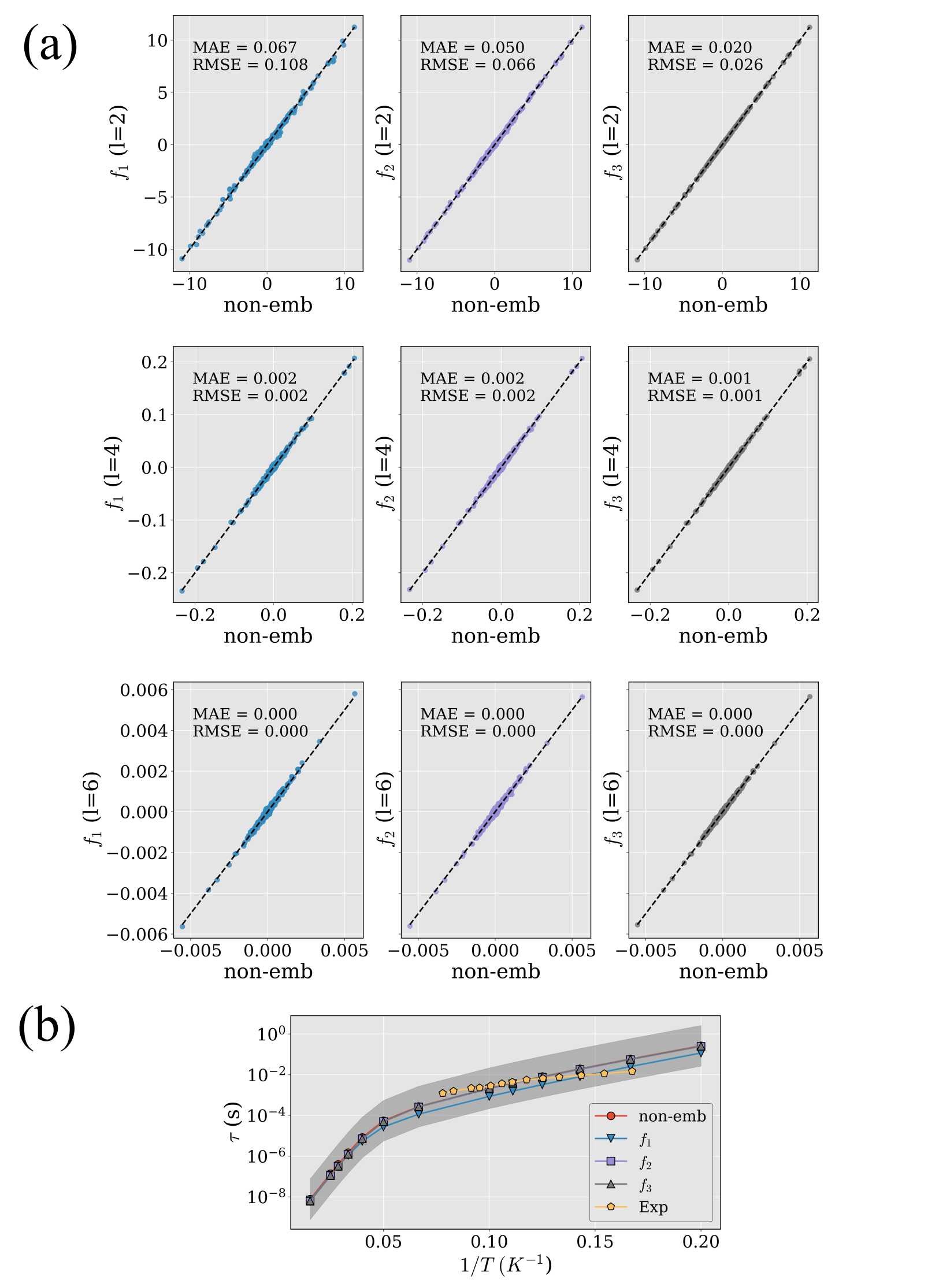}
    \caption{(a) The parity plots for the crystal field parameter derivatives (k=2,4,6) corresponding to three DMET fragments $f_1,f_2,f_3$ against the non-embedded CASSCF, for Complex IV. (b) The total relaxation times ($\tau$) as a function of temperature $(T)$ for Complex IV. The dark gray region represents accuracy region of an order of magnitude around the pristine CASSCF relaxation times. The experimental data is obtained from Ref. \citenum{liu2016stable}.}
    \label{fig:dy-9-dmet}
\end{figure}

For Dy(III), the crystal-field Hamiltonian formally contains contributions up to $l=14$ (\Cref{eqn:tesseral}). Since the contributions from terms beyond $l=6$ are negligible, only the $l=2$, $4$, and $6$ CFP derivatives were included. The corresponding parity plots are shown in \Cref{fig:dy-9-dmet}(a). All three fragmentations reproduce the non-embedded CFP derivatives well, with only minor differences among the fragments.

The relaxation times obtained from the embedded calculations are shown in \Cref{fig:dy-9-dmet}(b). All three fragments reproduce the non-embedded relaxation profile across the full temperature range. The differences observed for $f_1$ are small and remain well below one order of magnitude. As in the KD analysis, the remaining discrepancies can be traced to differences in the higher excited KDs rather than the lowest states.

The deviation from experiment below approximately $40~\mathrm{K}$ has also been reported previously.~\cite{mondal2022unraveling,haldar2023local} This discrepancy has been attributed to the absence of acoustic and Brillouin-zone phonons in the model. Nevertheless, both the embedded and non-embedded calculations remain within one order of magnitude of the experimental relaxation times.

\paragraph{Complex-V:} Complex-V is a Dy(III)-based single-molecule magnet with a coordination environment distinct from that of Complex-IV. In contrast to the connected \ce{H2bbpen} ligand framework, this complex contains water molecules in the equatorial plane and a \ce{^tBuPO(NH^iPr)_2} axial ligand. The relaxation-rate calculations were performed using the first-coordination-shell optimized geometry reported in Ref.~\citenum{mondal2022unraveling}.

\begin{table}[H]
    \centering
    \begin{tabular}{ccccc}
    \hline
         States& CASSCF & $f_1$ & $f_2$ & $f_3$\\
         \hline
        $KD_0$ & 0 & 0 & 0 & 0\\
        $KD_1$ & 426 & 443 & 440 & 431\\
        $KD_2$ & 759 & 795 & 789 & 769\\
        $KD_3$ & 914 & 994 & 979 & 935\\
        $KD_4$ & 956 & 1016& 1006 & 973\\
        $KD_5$ & 995 & 1065 & 1052 & 1014\\
        $KD_6$ & 1041 & 1096 & 1089 & 1058\\
        $KD_7$ & 1099 &  1169& 1156 & 1118\\
        \hline
    \end{tabular}
    \caption{Energies of the lowest Kramers' doublets (in $\mathrm{cm}^{-1}$) for Complex-V using CASSCF and various CAS-DMET fragmentations.}
    \label{tab:lowestkd5}
\end{table}

The energies of the eight KDs belonging to the ground-state $J=15/2$ manifold are reported in \Cref{tab:lowestkd5}. The excited KD energies show a slower convergence with fragment size than in Complex-IV, although the deviations decrease systematically from $f_1$ to $f_3$.

\begin{figure}[H]
    \includegraphics[width=0.9\linewidth]{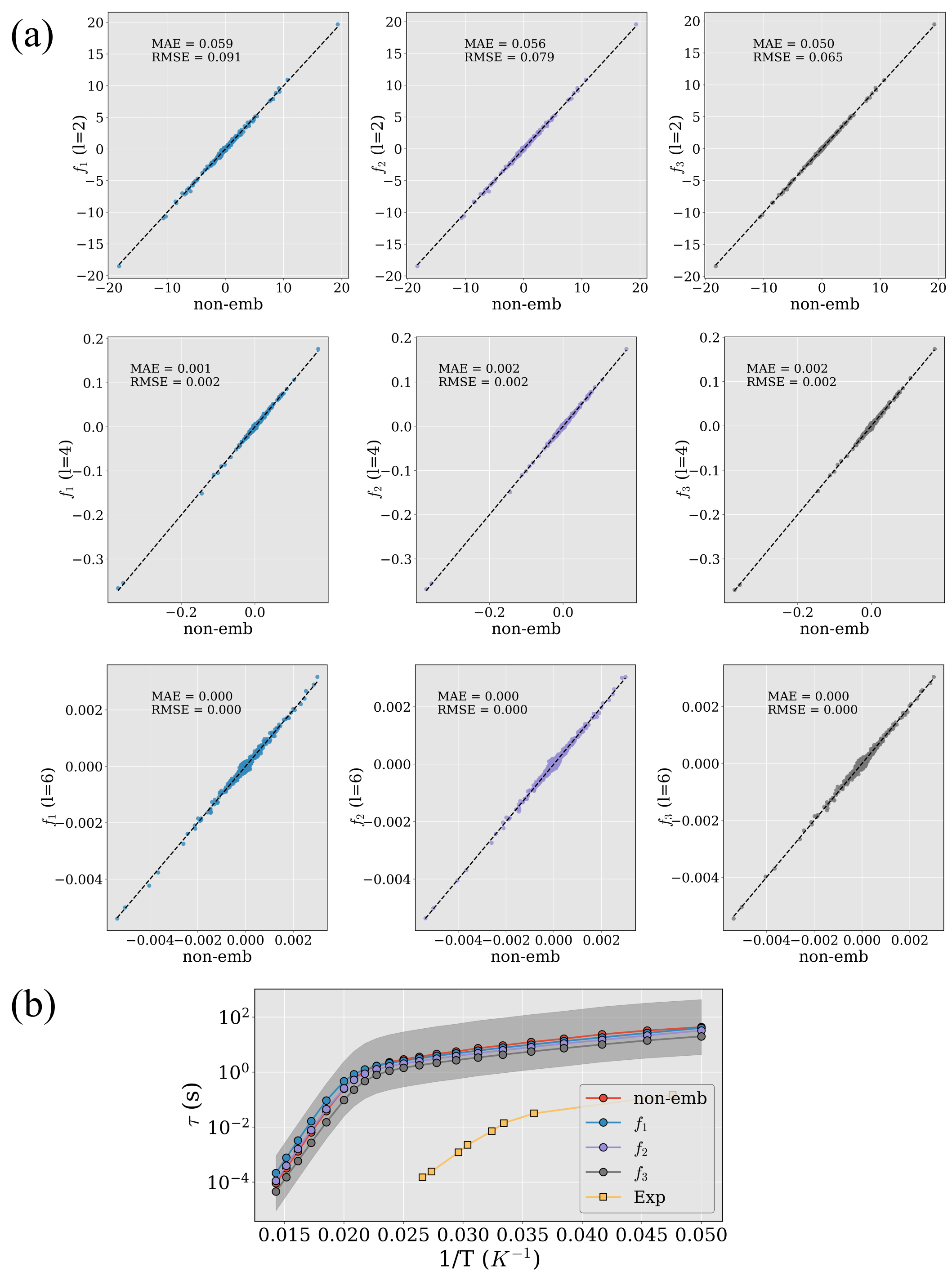}
    \caption{(a) Comparison between the crystal field parameter derivatives (k=2,4,6) corresponding to three DMET fragments $f_1,f_2,f_3$ against the non-embedded CASSCF, for Complex V. (b) The total relaxation times ($\tau$) as a function of temperature $(T)$ for Complex V. The dark gray region represents accuracy region of an order of magnitude around the pristine CASSCF relaxation times. The experimental data is obtained from Ref. \citenum{gupta2016air}.}
    \label{fig:dy-7-dmet}
\end{figure}

Despite these differences, the crystal-field parameter derivatives are reproduced well by all three fragmentations, as shown in \Cref{fig:dy-7-dmet}(a). Consequently, the relaxation times obtained with CAS-DMET closely follow the non-embedded CASSCF results across the full temperature range (\Cref{fig:dy-7-dmet}(b)). For all fragmentations, the relaxation times remain within the commonly accepted accuracy of one order of magnitude relative to the non-embedded reference.

Both the embedded and non-embedded calculations deviate significantly from experiment. This behavior was previously reported in Ref.~\citenum{mondal2022unraveling} and attributed primarily to structural differences between the DFT-optimized and experimental geometries. In particular, one of the \ce{Dy-OH2} bond distances is approximately 0.07~\AA\ longer in the X-ray structure compared to the optimized distance(see Supporting Information, Figure~S4). We expect that the effect of phonon q-points can also play an important role and possibly improve the agreement with experiment. Since the purpose of the present work is to assess the ability of CAS-DMET to reproduce the corresponding non-embedded multireference results, the agreement obtained across all fragmentations demonstrates that the embedding procedure does not introduce additional errors in the calculated spin-phonon relaxation rates. 
\pagebreak

\subsection{Molecular Crystal}
To assess the effect of the crystalline environment on spin relaxation, we performed periodic CAS-DMET calculations for Complex-I using the experimental molecular crystal structure.
These calculations are not feasible with conventional periodic CASSCF, but become accessible within the embedding framework.
Only the smallest fragment, $f_1$, consisting of the \ce{CoS4} unit (\Cref{fig:frag_scheme}(b)), was considered.
The inclusion of the crystal environment consisting of acetonitrile and tetraphenyl phosphonium salt leads to only modest changes in the low-lying KD energies. As shown in \Cref{tab:lowestkd_crystal}, the first excited KD differs by approximately $9~\mathrm{cm}^{-1}$ from the isolated-molecule CASSCF result, indicating that the low-energy spin-orbit manifold is largely determined by the local coordination environment of the Co center.

\begin{table}
    \centering
    \begin{tabular}{ccccc}
    \hline
         States& CASSCF (Complex-I) & $f_1$ (Crystal-I)\\
         \hline
        $KD_0$ & 0 & 0 \\
        $KD_1$ & 279 & 270\\
        \hline
    \end{tabular}
    \caption{Energies of the lowest Kramers' doublets (in $\mathrm{cm}^{-1}$ for Complex-I (using CASSCF) and Crystal-I (using CAS-DMET).)}
    \label{tab:lowestkd_crystal}
\end{table}

\begin{figure}[H]
    \centering
    \includegraphics[width=0.6\linewidth]{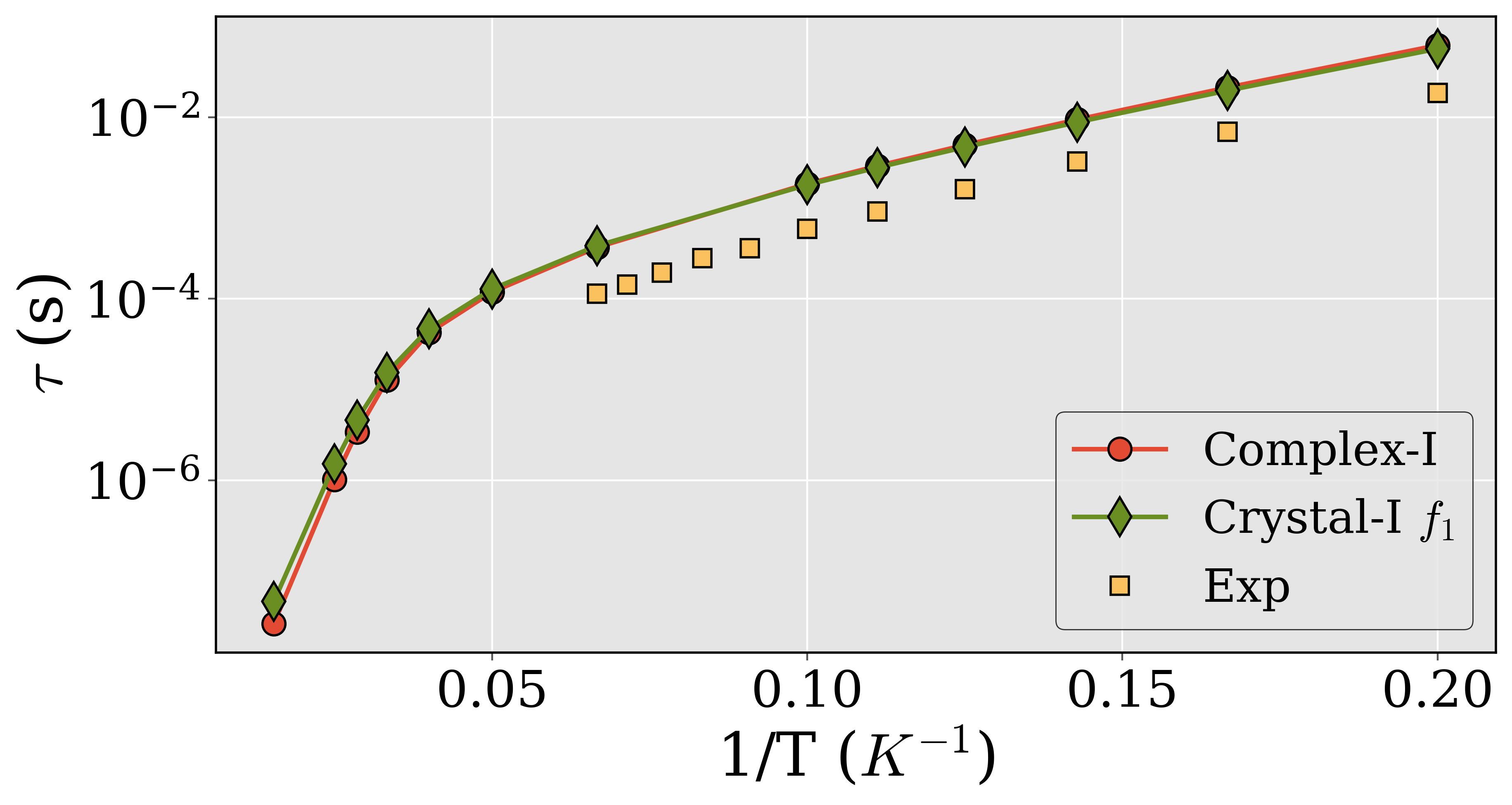}
    \caption{The relaxation times ($\tau$) for the full molecular crystal of Complex-I with the first coordination shell of the Co complex as the DMET fragment. The $\tau$ for the molecular Complex-I (only the complex without any solvent and without periodic boundary conditions) is also shown.}
    \label{fig:crystal}
\end{figure}

Using finite-difference displacements of $\pm0.01$~\AA, the periodic CAS-DMET calculations yield smooth crystal-field parameter derivatives and relaxation times that closely match both the isolated-molecule calculations and experiment (\Cref{fig:crystal}). Thus, although embedding makes the explicit treatment of the crystal environment computationally accessible, the present calculations show that the extended molecular crystal has only a minor effect on the spin-phonon relaxation dynamics of Complex-I. Instead, the relaxation mechanism is largely local, with the first coordination sphere capturing the dominant physics.

These calculations demonstrate that CAS-DMET provides a practical route for incorporating periodic crystalline environments into multireference spin-phonon simulations. In the present case, a fragment containing only the first coordination shell of the metal center is sufficient to reproduce the experimental relaxation behavior while retaining the atomistic description of the crystal.

\section{Discussion and Conclusions}
We have presented an approach for the calculation of spin-phonon relaxation rates in molecular and crystalline systems based on embedding a CASSCF wave function in a mean field environment.

The method was applied to a series of cobalt- and dysprosium-based single-molecule magnets spanning different coordination environments. Across all systems studied, fragments containing only the first coordination shell of the magnetic center reproduced the non-embedded CASSCF relaxation rates within approximately one order of magnitude, while reducing the size of the correlated problem to $26-68\%$ across different systems. As expected, the agreement improved systematically as the fragment size increased.

The calculations show that relaxation rates are sensitive to changes in the crystal field parameters and high lying KDs as predicted by different fragmentation sizes. While KD energies converge with fragment size, errors of $\approx90~\rm{cm}^{-1}$ in the smallest fragment $f_1$ KDs reflect in Raman rates at low temperatures. Errors of more than $\approx20~\rm{cm}^{-1}$ in crystal-field parameter derivatives propagate directly into the relaxation times and can lead to deviations less than an order of magnitude, as observed in Complex-II. 

We also applied the approach to a molecular crystal using periodic CAS-DMET. For the cobalt system considered here, the relaxation rates obtained in the crystal are similar to those of the isolated molecule, suggesting that the dominant spin-phonon coupling mechanisms remain local in character. More importantly, the calculation demonstrates that spin-phonon relaxation rates can be computed in a periodic multireference framework using an impurity containing only $\approx10\%$ of the total basis functions.

The systems studied here are relatively weakly correlated and can still be treated with conventional CASSCF. The real advantage of embedding approaches will emerge for systems where a full multireference treatment is no longer feasible, such as magnetic impurities in covalent solids, molecular materials with large unit cells, and extended frameworks. In these cases, embedding may provide a practical route to spin-phonon calculations that would otherwise be inaccessible.

Future work will focus on extending the methodology to higher-level electronic structure approaches, including NEVPT2-DMET \citenum{mitra2022periodic} and DME-PDFT \citenum{mitra2023density,verma2025dmepdft}, as well as incorporating spin-spin interactions and periodic spin-orbit coupling effects.

\begin{acknowledgement}
We acknowledge the University of Chicago’s Research Computing Center for their computational resources. This material is based upon work supported by the U.S. Department of Energy, Office of Science, National Quantum Information Science Research Centers by the Midwest Integrated Center for Computational Materials (MICCoM) as part of the Computational Materials Sciences Program funded by the U.S. Department of Energy.
M.R.H. and L.G. are partially supported by the U.S. DOE Office of Science, Office of Basic Energy Sciences, Heavy Element Chemistry Program under Award Number DE-SC0022572. A.L. acknowledges the funding from the European Research Council (ERC) (grant agreement No. 948493).
\end{acknowledgement}

\begin{suppinfo}

\end{suppinfo}

\bibliography{ref}
\end{document}